\documentclass[12pt]{article}

\usepackage{graphicx} 
\usepackage{a4wide}
\usepackage{cancel}
\usepackage{amsmath}
\usepackage{comment} 
\usepackage{amsfonts}
\usepackage{graphics}
\usepackage{minitoc}
\usepackage{titletoc}
\usepackage[T1]{fontenc}
\usepackage[utf8]{inputenc}
\usepackage{rotating}

\usepackage{booktabs}
\usepackage[margin=1in]{geometry}
\usepackage{subfig}
\usepackage[toc,page]{appendix}
\usepackage{threeparttable}
\usepackage[resetlabels]{multibib}
\newcites{App}{References - Supplementary Materials}
\usepackage{color}
\usepackage{lscape}
\usepackage{bbold}
\usepackage{umoline}
\usepackage{fancyhdr}
\usepackage{authblk}
\usepackage{chngcntr}
\usepackage{etoolbox}
\usepackage{lipsum}
\usepackage{float}
\usepackage{afterpage}
\usepackage[round]{natbib}
\usepackage[section]{placeins}
\usepackage{mwe} 
\usepackage{caption}
\DeclareCaptionFormat{myformat}{\textsc{#1#2}#3\par}
\usepackage{multicol}
\usepackage{appendix}
\usepackage{amssymb}
\usepackage{tikz}
\usepackage{sgame}
\usepackage{tabularx}
\usepackage{multirow}
\usepackage[bottom,multiple]{footmisc}
\usepackage[table]{colortbl}
\usepackage{url}
\usepackage{setspace}
\usepackage{pdfpages}
\usepackage{chronosys}
\usepackage{xstring}

\usepackage{ragged2e}
\usepackage[hidelinks]{hyperref}
\hypersetup{
    colorlinks=true,
    citecolor=blue,
    linkcolor=blue,
    filecolor=blue,      
    urlcolor=blue,
}
\usepackage{cleveref}
\usepackage[T1]{fontenc}
\usepackage{newpxtext,newpxmath}

\newcites{Main}{References for Main Text}
\newcites{Supplementary}{References for Supplementary Materials}

\newcolumntype{Y}{>{\centering\arraybackslash}X}

\usetikzlibrary{decorations.pathreplacing, fit, calc}

\makeatletter
\def\ps@pprintTitle{%
  \let\@oddhead\@empty
  \let\@evenhead\@empty
  \def\@oddfoot{\reset@font\hfil\thepage\hfil}
  \let\@evenfoot\@oddfoot
}
\makeatother

\newtheorem{theorem}{Theorem}

\newtheorem{corollary}[theorem]{Corollary}

\newtheorem{proposition}{Proposition}

\renewcommand {\baselinestretch} {1.5} 
\title{\Huge\textbf{
A Criminology of Machines\thanks{I am grateful to Alberto Acerbi, Massimo Airoldi, Alberto Aziani, Gianmarco Daniele, Roberto Dessì, Simon Egbert, James Evans, Laura Ferrarotti, Gary LaFree, Joel Leibo, Bruno Lepri, Giada Pistilli, and Jacopo Staiano for their precious feedback and suggestions on earlier drafts of this work.}}}
 
\author{Gian Maria Campedelli\\
\normalsize
\textit{Fondazione Bruno Kessler}
 \date{}
}

\begin{document}

\maketitle
\begin{abstract}
\singlespacing
\textit{
While the possibility of reaching human-like Artificial Intelligence (AI) remains controversial, 
the likelihood that the future will be characterized by a society with a growing presence of autonomous machines is high. 
In fact, autonomous AI agents are already deployed and active across several industries and digital environments.  
This trajectory points to a progressive hybridization of society marked by new forms of social interaction at both micro and macro levels.
Alongside traditional human-human and human-machine interactions, machine-machine interactions are poised to become increasingly prevalent.
Given these developments, I argue that criminology must begin to address the implications of this transition for crime and social control. 
Drawing on Actor–Network Theory and Woolgar’s decades-old call for a sociology of machines — frameworks that acquire renewed relevance with the rise of AI foundation models and generative agents — I contend that criminologists should move beyond conceiving AI solely as a tool. 
Instead, AI agents should be recognized as entities with agency, understood as a multi-layered construct encompassing computational, social, and legal dimensions. 
Building on insights from the literature on AI safety, I thus examine the risks and challenges associated with the rise of multi-agent AI systems, proposing a dual taxonomy to characterize the channels through which interactions among AI agents may generate deviant, unlawful, or criminal outcomes. 
I then advance and discuss four key questions that warrant theoretical and empirical attention: 
(1) Can we assume that machines will simply mimic humans? 
(2) Will crime theories developed for humans hence suffice to explain deviant or criminal behaviors emerging from interactions between autonomous AI agents?  
(3) What types of criminal behaviors will be affected first? 
(4) How might this unprecedented societal shift impact policing?  
These questions form the core of this article, underscoring the urgent need for criminologists to theoretically and empirically engage with the implications of multi-agent AI systems for the study of crime and play a more active role in debates on AI safety and governance.
}
\end{abstract}

\thispagestyle{empty} 

\clearpage
\pagenumbering{arabic} 

\newgeometry{left=1in,right=1in,top=25mm,bottom=25mm}
\section{Introduction}\label{intro}

The possibility (and desirability) of reaching human-like AI\footnote{Or General Artificial Intelligence or even Superintelligence and AI Singularity, or whatever exotic name associated with AI becoming equally or more intelligent than humans.} remains highly debated. And so it has been since 1956, the year in which the Dartmouth Summer Research Project on Artificial Intelligence, the event symbolically marking the beginning of AI as a discipline, was held. Discussions and predictions about human-like AI have been revamped in recent years due to the explosion and diffusion of foundation models, and chiefly Large Language Models (LLMs) \citep{KimRoadArtificialSuperIntelligence2024, IshizakiLargelanguagemodels2025}.

Notwithstanding the actual reachability of human-like AI (or the time horizon associated with this scenario),\footnote{Admittedly, two topics the author of this piece has no sufficient knowledge to provide definitive answers about. For relevant surveys and reports scanning expert predictions about this very topic, see \cite{MullerFutureProgressArtificial2016c, GraceViewpointWhenWill2018, AssociationfortheAdvancementofArtificialIntelligenceAAAI2025Presidential2025}.} the world -- and therefore human society -- will soon witness an increasing presence of autonomous AI agents.\footnote{While many definitions exist I borrow the popular one proposed by \cite{WooldridgeAgenttheoriesarchitectures1995}, who wrote that an intelligent or AI agent is a software-based computer system that is characterized by a) autonomy, b) social ability, c) reactivity, and d) pro-activeness. Another broader definition, recently proposed by \cite{MitchellFullyAutonomousAI2025}, states that AI agents are \textit{``computer software systems capable of creating context-specific plans in non-deterministic environments''}.} Breakthroughs in intelligent systems have already led to the development and deployment of autonomous agents in different sectors and industries. Prominent examples include the military domain \citep{palantir2025aip}, finance and banking \citep{ParkEnhancingAnomalyDetection2024, BousquetteDigitalWorkersHave2025}, and logistics \citep{BensingerAmazonsdeliverylogistics2025}, with the prospect that autonomous AI agents will spread across more and more contexts (e.g., healthcare, see \cite{MoritzCoordinatedAIagents2025}).

These developments signal the rise of a hybrid society in which agency is no longer the exclusive prerogative of humans or animals.\footnote{Institutions and legal entities also exercise agency. However, they are not central to my argument here, since they can be understood, in a stylized way, as collectives of humans. They are founded and maintained by humans. My focus instead is on entities at the individual level -- ontologically, epistemologically, and phenomenologically distinct from humans. Machines fall into this category.} AI agents are acquiring capacities to perceive, decide, adapt, and engage socially. This hybridization introduces a novel typology of interactions. For most of history, interaction occurred primarily among biological entities; in recent decades, however, advances in robotics, computing, and especially social media have produced a second modality, centered on human–machine exchanges. This shift has already necessitated new research fields devoted to examining how we communicate, collaborate, and co-exist with technology \citep{Hochumanmachineinteraction2000, RahwanMachinebehaviour2019, Tsvetkovanewsociologyhumans2024b}.

\begin{figure}[!hbt]
    \centering
    \includegraphics[width=0.9\linewidth]{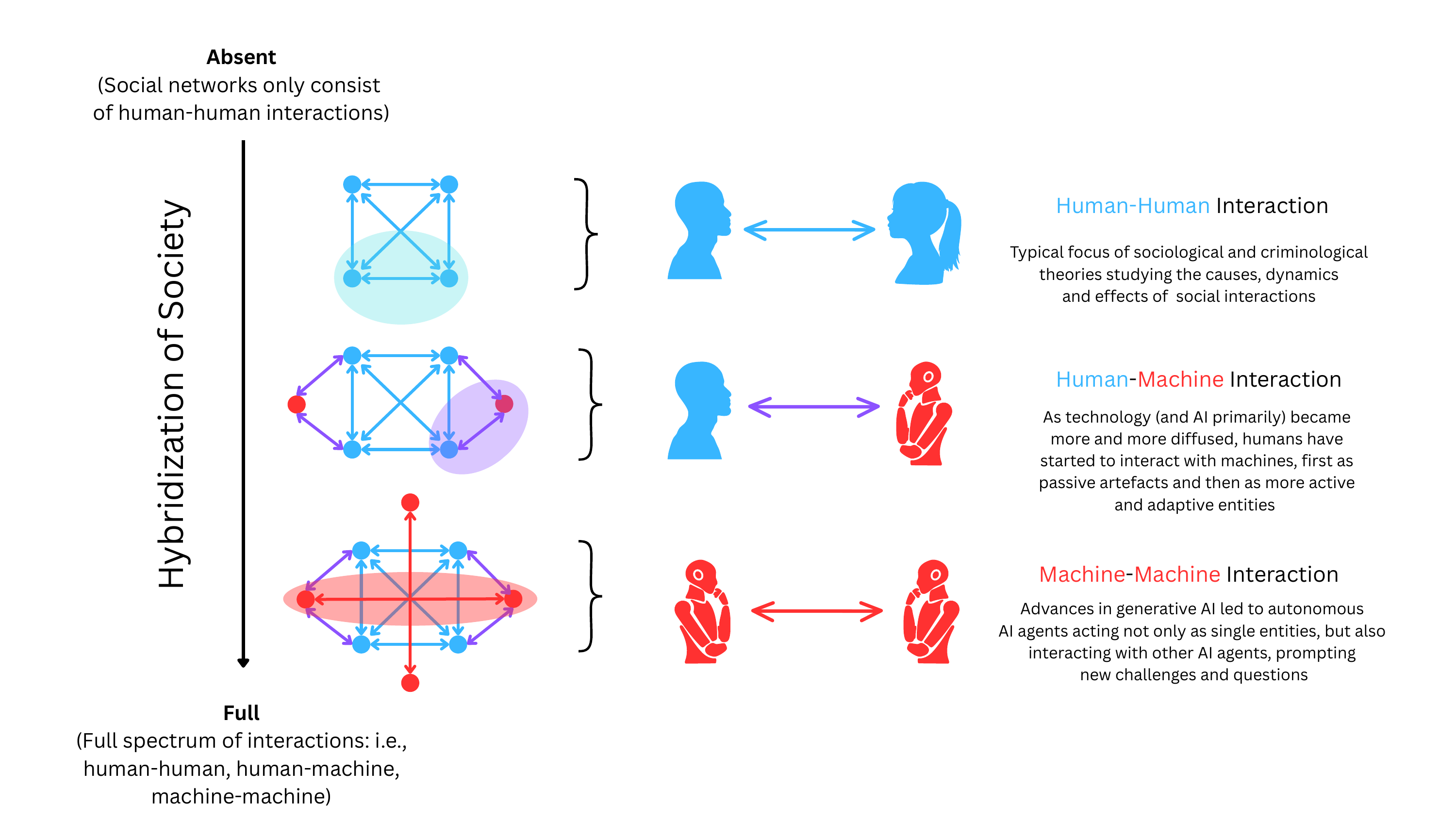}
    \caption{Stylized visualization depicting the ongoing process of hybridization of society. For most part of history, social networks only consisted of human (or biological) entities. Over the centuries, given technological advancements, humans have started to interact with machines, thus generating social networks that also included the human-machine dimension. Nowadays, we are witnessing a third phase characterized by an increasing autonomy of machines, and particularly AI agents, which are able to generate and maintain relationships with other AI agents, thus implying a third typology of interactions, i.e., the machine-machine one.}
    \label{fig:hybridsociety}
\end{figure}

Nowadays, we stand at the precipice of another paradigm shift, one that may possibly carry consequences of unprecedented scale. The rapid proliferation of truly autonomous (generative) AI agents\footnote{I refer to generative AI agents -- which are currently the state-of-the-art and may or may not in the future be surpassed by agents built on entirely different premises -- as agents powered by foundation models, such as (mostly) LLMs, Vision Foundation Models (VFMs), or Multimodal Models, such as GPT-4o \citep{OpenAIGPT4oSystemCard2024}.} marks the emergence of a third and distinct typology of interaction, i.e., the machine–machine one, a typology that for the first time does not entail any biological entity, one for which our almost complete ignorance may become hugely problematic and consequential (Figure \ref{fig:hybridsociety}).

This critical need for shedding light on machine–machine behavior is already resonating within the AI and computer science communities. Fueled by the widespread use and availability of LLMs, recent scholarship has investigated behavioral patterns of LLM-powered AI agents in different contexts \citep{DafoeOpenProblemsCooperative2020, LiuDynamicLLMPoweredAgent2024, DengAIAgentsThreat2025, LiSingleTurnSurveyMultiTurn2025, AsheryEmergentsocialconventions2025}. Whether motivated by the potential to simulate complex social phenomena or the desire to understand the emergent dynamics generated by conversations between LLMs, scholars have been attracted by the manifold questions that these new forms of interactions pose for scientific research. In this context, one of the aspects that is fostering notable discussions concerns the risks associated with multi-agent AI systems, i.e., systems of AI agents interacting with each other with no human mediation \citep{HammondMultiAgentRisksAdvanced2025, deWittOpenChallengesMultiAgent2025}.

Such a discussion is not only speculative and theoretical, but is already substantiated by empirical evidence of unintended deviant and unlawful behaviors by interactive AI agents both in research \citep{FishAlgorithmicCollusionLarge2024, CampedelliWantBreakFree2024a, BichlerAlgorithmicPricingAlgorithmic2025} as well as in real-world practical domains, as shown by scandals of collusion in algorithmic pricing \citep{PriluckWhenBotsCollude2015}. Multi-agent AI systems, in fact, introduce distinct risks by enabling agents to learn from, adapt to, and coordinate with one another in ways that are not always predictable or transparent. This interactive dynamic can give rise to emergent behaviors, patterns of action that are not explicitly programmed and may be difficult to detect, explain, or control. As a result, these systems can generate different types of harm, including fraud, manipulation, discrimination, or the dissemination of disinformation, sometimes absent any direct human intervention and possibly through novel decision-making or atypical behavioral patterns. These developments challenge traditional criminological categories and raise pressing questions about responsibility, regulation, and prevention in a world increasingly shaped by non-human actors.

In light of these developments, in this article, I argue about the necessity to engage with the prospect of a criminology of machines, i.e., a criminology that considers AI agents as social agents interacting with each other and that reason and discuss about the potential effects and implications that such agency and autonomy may have on criminal phenomena and policies and institutions aiming at preventing or controlling crime.

Inspired by previous theoretical conceptualizations by \cite{WoolgarWhynotSociology1985c} and champions of Actor-Network Theory \citep{Latouractornetworktheoryfew1996, LawActorNetworkTheory1999}, I contend that we, as a scholarly community, should begin engaging with these foundational issues. I suggest that doing so opens the door to a series of further inquiries, which I will outline and explore in the remainder of this piece. Moreover, I argue that criminologists could contribute -- jointly with experts from the AI community -- to the efforts to predict, contain, mitigate, and govern the risks emerging from interactive AI agents.

The article is structured as follows: In the next section, I will briefly discuss how crime and AI have been traditionally studied together, calling for a paradigm shift that moves from AI as a tool to the recognition of AI agents as an active part of society. In doing so, I draw on sociological theories that conceptualize non-human entities as central to the understanding of society, highlighting how advances in AI make such a framework particularly appealing for re-evaluating the role of intelligent machines in our world. Furthermore, taking inspiration from recent work in philosophy, I propose a definition of AI agency encompassing three dimensions (i.e., computational, social, legal), aiming to formalize a conceptual platform that both describes the current state of AI agents and offers a lens for analytical and theoretical scrutiny. In the third section, I provide a concise overview of how AI agents are becoming increasingly autonomous and how scholars across disciplines have already started to reflect on the possible outcomes and implications of this process, highlighting potential risks associated with AI agents learning from each other, as well as discussing two channels through which multi-agent AI systems may lead to the commission of deviant, unlawful, or criminal behaviors. In the fourth section, I lay out four important questions we should carefully consider in our quest toward a criminology dedicated to machines. Before concluding the article, I also discuss the role criminologists should have at the beginning of this new era.

\section{AI and Crime: Shifting the Perspective}

\subsection{AI as a Tool}

\paragraph{A Tool for Research}
Despite the prevailing notion that the convergence of AI and criminology is a recent development, the relationship between these domains traces back to the 1980s \citep{CampedelliMachinelearningcriminology2022b}. Many things have changed since the first attempts to build programs able to predict crime events -- not so much in terms of goals, but in terms of popularity, computing power, computational architectures, and richer data availability. In the 1980s and 1990s, attempts at using AI to address crime-related problems were rare and relied on much less powerful hardware, often on symbolic architectures or expert systems (see \cite{IcoveAutomatedCrimeProfiling1986}, \cite{RatledgeHandbookArtificialIntelligence1989}, and \cite{HernandezArtificialintelligenceexpert1990}). More recently, the use of machine learning and deep learning has gained traction -- with an explosion of publications in the past five years, in criminology and computer science alike. Scholars can now process large amounts of data on personal laptops (or, in the most demanding cases, via cheap cloud servers) and perform prediction or forecasting tasks using more expressive, flexible methods, often based on tree-based or neural architectures.

Works exploiting these methods now appear not only in transdisciplinary journals or venues in computer science, such as AI conferences, but also in orthodox criminology journals, signaling the shift from unorthodox to mainstream methods.

This compact (and therefore not at all comprehensive) depiction of the current landscape demonstrates that AI in criminology -- and, more broadly, in the social sciences -- has been seen, studied, and utilized as a tool, a means to an end. In most cases, machine and deep learning algorithms are deployed to solve a specific task (such as forecasting recidivism, e.g., \cite{BerkCriminalJusticeForecasts2012, Dresselaccuracyfairnesslimits2018d}, or predictive policing, e.g., \cite{ferguson2016policing}, \cite{kaufmann2019predictive}), to test theories \citep{MolinaMachineLearningSociology2019c}, or, less commonly, to discover hypotheses in the attempt to generate new research questions in an \textit{agnostic} fashion, as proposed by \cite{GrimmerMachineLearningSocial2021}. In substance, most criminologists and social scientists see AI as a competitor of traditional statistical methods -- a set of techniques that make the quantitative researcher's job easier.\footnote{Importantly, the use of AI methods to address criminological research questions applies not only to machine and deep learning approaches but also to LLMs. See, for instance, \cite{AdamsNomanshand2024} and \cite{RelinsUsingInstructionTunedLarge2025}.}


\paragraph{A Tool for Committing Crime}
AI, unfortunately, is not only seen and utilized as a flexible and powerful tool for research. It is also exploited as a technology for committing crime \citep{CaldwellAIenabledfuturecrime2020, BlauthArtificialIntelligenceCrime2022}. \cite{KingArtificialIntelligenceCrime2020} introduced the term Artificial Intelligence Crime (AIC) to describe the use of AI for unlawful purposes, a phenomenon studied across multiple disciplines. In their seminal paper, they examine three key questions: who should be considered the true perpetrator of an AI-enabled crime (a human or the artificial agent itself?), how an AIC should be defined, and in what ways such crimes are typically carried out.

Building on this, \cite{HaywardArtificialintelligencecrime2021a} classify AIC into three subcategories: (1) crimes \textit{with} AI, (2) crimes \textit{against} AI, and (3) crimes \textit{by} AI. The first, and arguably most common, refers to cases where AI is deployed as a tool for malicious purposes, amplifying existing criminal threats and generating new risks. Examples include AI-powered drones used for targeted killings and AI-driven social engineering attacks in cyberspace.

Crimes against AI involve exploiting vulnerabilities in AI systems. Such acts include corrupting training data or launching adversarial attacks which can produce unintended or unlawful outcomes.

The third category, crimes by AI, encompasses cases where AI operates as an intermediary in unlawful activity. Here, AI’s growing autonomy and capacity for specialized tasks enable it to deviate from deterministic behaviors. Examples include experimental cases of market manipulation and collusion \citep{Martinez-MirandaLearningunfairtrading2016, EzrachiTwoArtificialNeural2017}, as well as real-world incidents such as an AI agent purchasing illegal goods online \citep{KasperkevicSwisspolicerelease2015}. According to \cite{HaywardArtificialintelligencecrime2021a}, this subcategory raises critical questions of liability and agency -- issues I will return to later in this manuscript.

While these categorizations are useful, much of the literature portrays AI primarily as a tool for unlawful acts, with humans as the central orchestrators and beneficiaries. This perspective, however, only partially reflects the current landscape. As AI capabilities advance, there is a growing need for a more comprehensive framework that reconsiders the role of AI agents in society and their potential involvement in criminal behavior.

\subsection{AI and Crime: Shifting the Perspective}

\subsubsection{AI Agents as an Integral Part of Society}
While I advocate the use of methods ported from the AI community to study crime, and while I recognize the relevance of studying and countering the use of AI as a tool for committing crimes, I argue that it is time for criminologists to adopt a substantial shift of perspective. Today, AI is not confined to models and algorithms that solve criminology-related tasks, nor should it be seen merely as a powerful technology in the hands of humans to perpetrate deviant, unlawful, or criminal behaviors.\footnote{Three overlapping but distinct terms will be used throughout the paper to describe harmful or disruptive behaviors that may emerge from machine–machine interactions: \textit{deviant behaviors}, \textit{unlawful behaviors}, and \textit{criminal behaviors}.

\textit{Deviant behaviors} refer to actions by artificial agents that diverge from established technical, social, or normative expectations, even if they do not violate formal rules. In this sense, deviance is understood relative to norms of proper functioning, including safety protocols, ethical guidelines, or user expectations. For example, two AI agents colluding to manipulate an online marketplace in ways that distort prices, without explicit illegality, would constitute deviance.

\textit{Unlawful behaviors} designate actions by AI agents that contravene codified rules or regulations, irrespective of whether those actions would traditionally be classified as crimes. These include violations of civil law, contractual agreements, or regulatory mandates. For instance, AI agents that systematically breach intellectual property protections or privacy regulations would be considered unlawful.

\textit{Criminal behaviors} are a narrower subset, referring specifically to machine-driven acts that fall under criminal law, as defined by legislatures and enforced by courts. This category encompasses conduct that is explicitly prohibited and subject to penal sanctions -- for example, AI-enabled fraud, unauthorized system intrusions, or, in more extreme cases, physical harm facilitated by embodied AI systems.

This tripartite distinction is useful because it prevents premature conflation: not all deviance is unlawful, and not all unlawful conduct rises to the level of crime. Yet for criminological analysis, each layer matters. Deviant patterns may signal vulnerabilities before they escalate into unlawful or criminal acts, while unlawful but non-criminal violations may nonetheless destabilize social trust and institutional order.}

Contemporary AI agents are completely different entities compared to standard Random Forests or Support Vector Machines: the scope of generative AI agents is much broader, characterized by a more diverse set of capabilities and constrained by larger development costs. Notably, all works cited in the previous subsection regarding the use of AI tools for committing crime were published at least four years ago and focused on reinforcement learning approaches rather than generative AI (see Section \ref{sec:autonomousagentstoday} for a discussion of the differences between these two technologies). By contrast, agents powered through generative foundation models emerged recently\footnote{A word such as \textit{recently} has wildly different meanings when comparing the fields of AI and criminology. In the former, the pace of innovation and the sheer volume of publications imply that, in some cases and subfields, work published five years ago is already fatally outdated. In the latter, however, \textit{recently} may still apply to works published a decade ago or even earlier. I will not elaborate further on this discrepancy, but I am convinced it is related to the broader narrative of this work -- namely, the need for criminology and the social sciences to seriously consider how technological breakthroughs may generate societal consequences at a much faster pace than criminology has traditionally accounted for. This point is discussed in detail by \cite{TopalliFutureCrimeHow2020}.} and can communicate, plan, and perceive the environment, solving a multitude of general or specialized tasks with greater speed and versatility than before.\footnote{Relevant disclaimer: I am not blind to the many shortcomings of contemporary LLMs, exemplified by (often spectacular) hallucinations and their inability to solve extremely easy problems \citep{WilliamsEasyProblemsThat2024, XuHallucinationInevitableInnate2025, MalekFrontierLLMsStill2025}. LLMs (and foundation models in general) have many, clear limits. My argument is not that AI agents are more intelligent than humans; the argument is that they have reached a level of autonomy that allows them to act in interactive environments, that this new collective paradigm requires scholarly attention, and that their failures and hallucinations add a further layer of complexity to understanding and predicting their behaviors. Notably, the argument of this paper is not necessarily tied to generative AI: it would remain relevant even if, in the near future, other technological advancements surpass transformer-based architectures such as LLMs in their cognitive, reasoning, and operational capabilities.}

In light of this, social scientists -- and criminologists in particular -- should recognize AI agents as an integral part of society, if not in the present, then in a highly likely future. Given the increasing diffusion of autonomous AI agents, and given their growing ability to interact with each other, we should avoid seeing AI solely as a static toolbox: these agents will play an increasingly active role in shaping human everyday life and are therefore worthy of theoretical and empirical attention.

\subsubsection{Theoretical Premises}

\paragraph{Actor-Network Theory and Its Relevance for Multi-agent AI Systems.} This call to recognize AI agents as integral social entities is grounded in the fundamental principles of Actor-Network Theory (ANT) \citep{Latouractornetworktheoryfew1996, LawActorNetworkTheory1999, LatourReassemblingSocialIntroduction2007}. ANT offers a critical conceptual lens for criminology because it radically flattens the ontological hierarchy between humans and non-humans, which is now essential for understanding the increasing autonomy of AI systems and agents.
At its core, ANT conceptualizes all entities -- human or non-human, animate or inanimate -- as actants of equal analytical importance in the study of society. This perspective deliberately moves away from the assumption that human agency is privileged over the agency of things, including machines. Importantly, and despite the reference to technologies that were very distant from the ones I discuss in this paper, ANT has already been applied in the literature as a social constructivist platform to study and theorize technological advancements and their impact for criminology \citep{RobertActornetworktheorycrime2016}. \cite{BrowncriminologyhybridsRethinking2006}, for instance, argued against a simple binarization separating the human and the artificial, advocating for the use of ANT and the necessity to blend social theory with information theory to really comprehend contemporary criminal phenomena. Aligning with this argument, \cite{vanderWagenCybercrimeCyborgCrime2015} studies bot nets, i.e., networks of infected computers controlled by a user, building on the prescriptions of ANT, defining them as hybrid criminal actor-networks, underscoring its relevance to illuminate offending dynamics, victimization as well as countering approaches.

\paragraph{Symmetry, Mediation, and Translation.} Three dimensions of ANT are particularly relevant to the study of modern AI agents, especially when considering machine-machine interactions. First, the \textit{Generalized Postulate of Symmetry} insists on treating human and non-human actors symmetrically when explaining how associations and social order -- including illicit orders -- are constructed. Latour and colleagues argue that scholarly focus should rest on relationships and associations -- the "network" -- and on the ability of actants, regardless of their nature (human, algorithm, or infrastructure), to influence the creation or diffusion of these relationships. This is crucial for criminology, as a deviant outcome emerging from autonomous interactions between AI agents is fundamentally a function of the entire socio-technical network, not a mere consequence of human programming or intent alone.

Second and third, ANT emphasizes \textit{mediation} and \textit{translation}. ANT specifies that non-human entities are not simple passive tools, but active mediators that transform or reshape human intentions through their structure, constraints, and operational logic. Translation refers to the processes by which various actants align their interests, negotiate roles, and stabilize networks. In the context of multiple autonomous generative AI agents interacting -- a scenario where decisions and outputs recursively feed into other agents -- this mediation is powerful. The collective system can move into a "self-referential regime," where the network’s internal dynamics (such as synthetic-data drift) generate systemic deviations from human-like behavior, leading to outcomes that are entirely emergent and non-human. ANT, therefore, provides the necessary vocabulary to analyze crime not as a function of individual human intent, but as an emergent property of a dynamic, relational socio-technical system.

\paragraph{Revamping Woolgar's Call.} This theoretical perspective requires criminology (and, relatedly, sociology) to re-evaluate the importance of the non-human, echoing the decades-long appeal of \cite{WoolgarWhynotSociology1985c}. In his seminal work, Woolgar called for a sociology of machines with two specific goals. The first, largely pursued within Science and Technology Studies (STS), concerned the analysis of daily routines and narratives of the AI community and later spurred ethnographic work on algorithmic systems \citep{SeaverAlgorithmsculturetactics2017, CellardAlgorithmsfigurespostdigital2022, Christinethnographeralgorithmblack2020}, including in criminal justice settings \citep{BrayneTechnologiesCrimePrediction2021}. The second goal -- less developed but now critical -- was precisely to make intelligent machines the \textit{actual subject of sociological analysis}, challenging the idea that the social is a distinctly human category. As noted by \cite{AiroldiMachineHabitusSociology2021}, forty years later Woolgar’s argument is more relevant than ever due to the operational reality of machine agency. Advancements in AI, championed by LLMs and foundation models, make AI agents -- technological products equipped with unprecedented computational power and task-solving abilities -- available at scale.\footnote{Which means, also, that they are available to scholars outside the traditional communities that for decades worked on multi-agent AI systems (see \cite{TanMultiagentreinforcementlearning1993, FerberMultiAgentSystemsIntroduction1999, ShohamIfmultiagentlearning2007, SandholmPerspectivesmultiagentlearning2007}), thus enabling broader and more diversified analyses.} This technological shift means that what was only possible through abstract theorizing decades ago becomes operationally viable and empirically necessary for criminology today. Scholars can now design and observe the emergent properties of machine-machine networks to anticipate, diagnose or control potential emergent criminal phenomena. Criminologists have not yet ventured into this unexplored path. Yet, scholars in other fields have. Section \ref{galaxy} elaborates on relevant scholarship emerging from the social and computer sciences, with a specific focus on safety. Before that, however, the next subsection provides an operational definition of AI agency, crucial to conceptualize -- theoretically and analytically -- the entities that are the object of this article.

\subsection{A Multi-dimensional Definition of AI Agency}

In the subsection above I have mentioned several times the word \textit{agency}, to summarize the key messages of the theoretical works of Woolgar and champions of ANT, in an effort to delineate the need to open criminology to the machine dimension, that is, recognizing the role that AI agents will increasingly have as active entities in our society. However, I have not yet explicitly defined what \textit{agency} shall mean when referred to AI agents. This very topic is -- and has been -- the core focus of a vast scholarship that has gained even more prominence in recent years with the advent of generative AI. This scholarship entails two different traditions. The dominant standard view ties agency to internal mental states such as beliefs and desires, thereby implying that AI agents do not possess any agency \citep{FritzMoralagencyresponsibility2020, SwanepoelArtificialIntelligenceAgency2024}. By contrast, the non-standard view suggests that agency should be evaluated in terms of three fundamental criteria, namely observable interactivity, autonomy, and adaptability, treating the concept as a spectrum rather than a binary property \citep{FloridiMoralityArtificialAgents2004, DungUnderstandingArtificialAgency2025}. This perspective has gained traction as AI systems increasingly make consequential decisions in domains such as policing or healthcare, and it is the one I subscribe to. In fact, the advancements in AI agents and their massive diffusion across domains and industry, the impressive capabilities that foundational models demonstrate across tasks and skills, and the increasing development of multi-agent systems are the empirical demonstration of the existence of the three abovementioned fundamental criteria that delineate and qualify agency.

\begin{figure}
    \centering
    \includegraphics[width=0.95\linewidth]{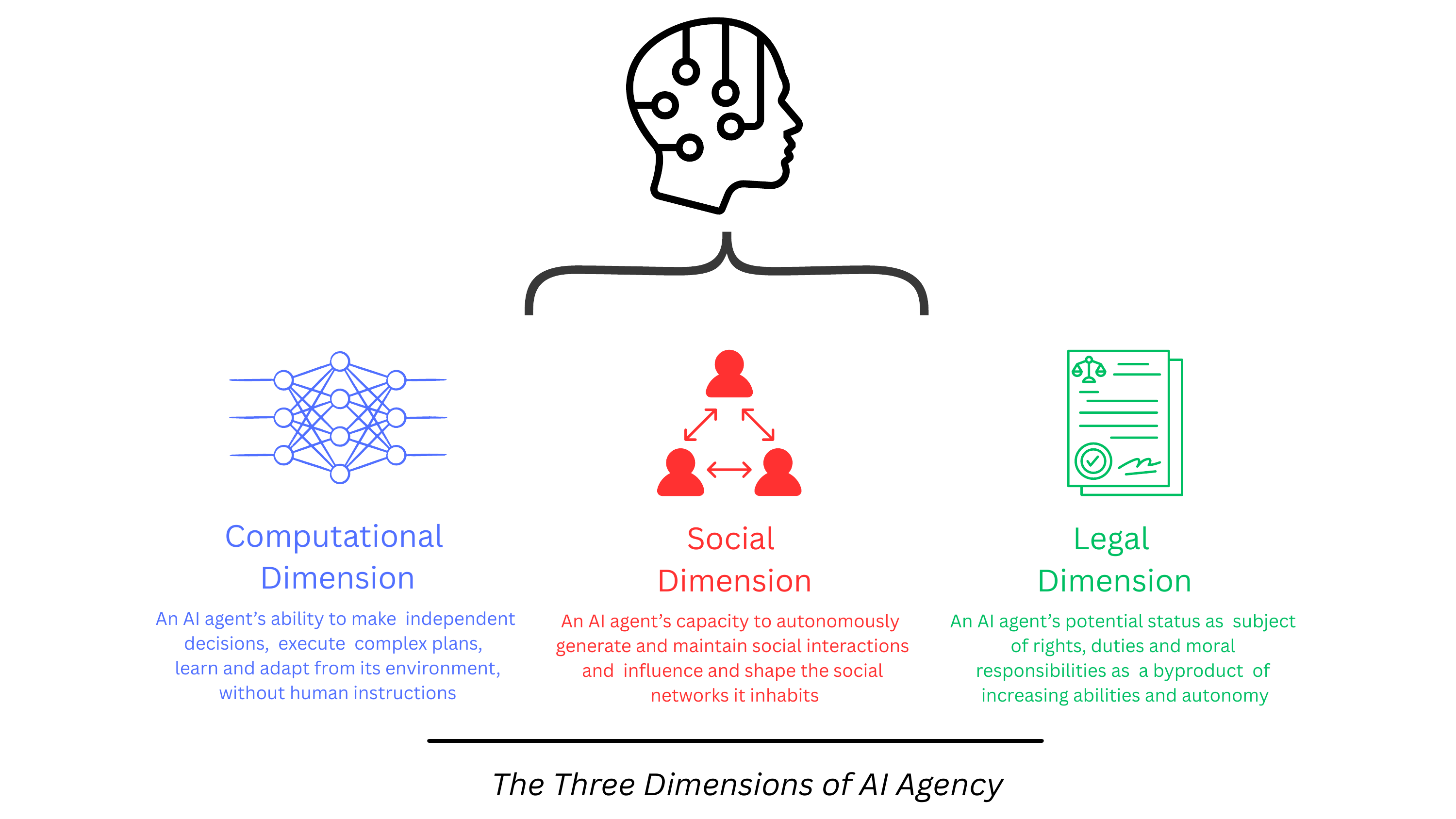}
    \caption{The three dimensions characterizing AI agency of modern, generative AI agents: Computational, Social, and Legal. By becoming more powerful and capable from a computational point of view, AI agents have also acquired increasing autonomy in their decision-making and, in turn, in their ability to interact with other agents. This increased autonomy poses potential issues from the legal standpoint.}
    \label{fig:placeholder}
\end{figure}

Within this latter tradition, \cite{FloridiAIAgencyIntelligence2025} recently proposed the terms \textit{Artificial Agency} and \textit{Artificial Social Agency} to define this specific new typology of agency that make AI agents distinct from biological purposefulness, mechanical determinism, and human intentionality. I agree that Artificial Agency differs from other categories scholars have studied for centuries (if not millennia) both in its individual and social forms, and I therefore argue that the computational dimension, which serves as the substrate for goal-directedness, is not sufficient to fully capture the nuances of agency in AI agents. Therefore, I draw inspiration from Floridi's taxonomy and propose below to consider AI agency as a multi-dimensional concept encompassing three interconnected dimensions: computational, social, and legal, which would serve as theoretical and analytical lenses to better understand what this \textit{machine dimension} operationally encompasses (Figure \ref{fig:placeholder}).

\paragraph{The Computational Dimension of AI Agency.} First, Computational Agency refers to the technical foundation of an AI's autonomy. This dimension describes the internal capacity of an AI to make independent decisions, execute complex plans, and learn from its environment without continuous, direct human instruction \citep{BurrellHowmachinethinks2016, BorchMachinelearningsocial2022}. This aspect becomes more salient today as it distinguishes modern generative AI agents from earlier, more deterministic models. The computational dimension almost perfectly overlaps with the elements in the definition of Artificial Agency provided by \cite{FloridiAIAgencyIntelligence2025}, as it focuses on machines' ability to solve extremely complex and specialized tasks in extremely short time horizons, operating at massive, distributed scales, and even misaligning with human goals which, according to some, should be constitutive of agency in AI (see, for instance, \cite{PopaHumanGoalsAre2021}). It follows that understanding the computational dimension of AI agency is critical for anticipating how the actions of a machine -- including potentially harmful or unlawful ones -- can emerge from the statistical decision-making processes that govern its functioning, creating new challenges for policing and forensics. Yet, only focusing on this dimension underplays the fundamental shift occurring in our society and overlooks the interactive autonomy of contemporary multi-agent AI systems.

\paragraph{The Social Dimension of AI Agency.} Therefore, the social dimension refers to the capacity of an AI agent to influence and shape the environment and social networks it inhabits. This dimension does not imply the existence of consciousness or intent or, more in general, internal mental states, hence refusing attitudes toward anthropomorphization of AI agents.\footnote{On the fallacies and problems associated with anthropomorphizing AI agents and algorithms, see \cite{WatsonRhetoricRealityAnthropomorphism2019} and \cite{PlacaniAnthropomorphismAIhype2024}.} Instead, it simply regards an actor's ability to produce tangible effects and alter relationships within a socio-technical system.\footnote{This view also broadly aligns with the so-called \textit{cultural} perspective on AI agency proposed by \cite{AiroldiMachineHabitusSociology2021}} This dimension lies fundamentally at the core of the narrative of the present work: AI agents are different from humans, yet they are becoming more and more autonomous, with little or no supervision from humans themselves, and this autonomy implies the ability to create relationships, both with humans and machines, hence certifying a social capacity \citep{RahwanMachinebehaviour2019, BorchMachinelearningsocial2022}. Such social capacity, in turn, presents a wide array of promises as well as, crucially, challenges that would entail criminal or deviant phenomena. In fact, interactive autonomy -- which represents the central feature defining the social dimension of Artificial Agency -- allows us to go beyond the perspective of AI agents acting solo, without being able to influence (or be influenced by) other machines, offering powerful lenses to possibly theorize and analyze multi-agent AI systems from a collective perspective.

\paragraph{The Legal Dimension of AI Agency.} Finally, the legal dimension pertains to an AI agent's potential status as a subject of rights, duties, and responsibilities. The legal dimension has been widely debated for decades \citep{KarnowLiabilityDistributedArtificial1996, HallevyCriminalLiabilityArtificial2010, ChestermanArtificialIntelligenceLimits2020}, with practical implications for regulation in recent years. While AI currently lacks legal personhood, the growing social and computational agency of these systems creates a significant criminological problem. Specifically, the increasing autonomy of AI agents -- also in relation to their social dimension -- gives rise to a potential "liability gap,"\footnote{Also known as "responsibility gap."} where it becomes increasingly difficult to assign blame and responsibility for a harmful or criminal act back to a human user, owner, or programmer \citep{MatthiasresponsibilitygapAscribing2004, SantonideSioFourResponsibilityGaps2021, FloridiDistributedMoralityInformation2017}. The challenges associated with this legal dimension thus highlight the urgent need for criminologists to engage with scientists involved in the design and development of multi-agent AI systems, as well as policy-makers and legal scholars, in order to discuss how the transition from single AI agents to collective AI behavior may require new frameworks and policies to be truly fair and effective.

\section{The Rise of Contemporary Multi-Agent AI Systems}\label{galaxy}

\subsection{Autonomous AI Agents Today
}\label{sec:autonomousagentstoday}

Recent years have seen the widespread diffusion of robots and AI agents across many sectors, impacting our daily lives. From autonomous vehicles to healthcare, from logistics to customer support, companies and organizations are increasingly leveraging advances in AI research to optimize their pipelines, carry out complex tasks, and improve efficiency. These agents typically operate independently, with varying levels of human supervision.

More recently, however, generative AI agents have expanded their autonomy by interacting with other AI agents, effectively leading to multi-agent systems. Drawing from \cite{deWittOpenChallengesMultiAgent2025}, a multi-agent system can be defined as a network of two or more autonomous AI agents characterized by six fundamental features: a) independent decision-making capabilities, b) ability to maintain private information, c) mutual interaction via communication channels or by modifying shared environments, d) a degree of autonomy, e) capacity to pursue their own objectives (or those delegated by human or artificial principals), and f) ability to adapt their behavior in response to external shocks.

As reported by \cite{HammondMultiAgentRisksAdvanced2025}, interactive AI agents are already deployed in finance and the military sector \citep{amplify2025aieq, palantir2025aip}, with the near-future prospect of becoming central in other areas such as health \citep{MoritzCoordinatedAIagents2025} and energy management \citep{CamachoLeveragingArtificialIntelligence2024,mayorkas2024ai}.

De Witt notes that contemporary multi-agent systems differ significantly from traditional ones (see \cite{WooldridgeAgenttheoriesarchitectures1995}) because they are powered by foundation models, such as LLMs, which provide flexible decision-making, communication, and generalizable reasoning capacity. Before the introduction of LLMs, multi-agent learning systems were primarily studied in the Reinforcement Learning community, particularly in Multi-Agent Reinforcement Learning (MARL) (see \cite{BusoniuComprehensiveSurveyMultiagent2008a} for a survey). A key difference between the two approaches is that, unlike MARL agents, generative AI agents possess knowledge and values acquired during pre-training and post-training, with social interaction occurring afterward. Interactive in-context learning thus appears as dynamic behavioral adaptation without parameter learning \citep{ChenPersonaVectorsMonitoring2025, DherinLearningtrainingimplicit2025}. In substance, generative agents are not \textit{tabula rasa} -- as agents in MARL are -- which raises challenges related to predictability in the interactive phase. These challenges are magnified by the lack of transparency regarding training data for closed-source models. Additionally, modern AI agents are \textit{generalist}, meaning they can communicate, perceive, and act in the environment. They are all built on transformer-based architectures, unlike the ad hoc, task-specific agents typical of the MARL literature.

\subsection{Safety of Multi-Agent AI Systems: A Brief Overview}

The study of AI agents -- and of multi-agent systems -- has a long tradition, but the advent of LLMs and other foundational models has spurred dramatic growth in scholarly attention to these areas. Foundation models offer new opportunities to study how AI agents interact and what dynamics characterize such interactions, depending on the context \citep{anthis2025llm}. Over the past three years, the literature has been flooded with works leveraging LLMs to address a range of questions, alongside the development of platforms for experimenting with multi-agent systems populated by LLM-powered agents (see, for instance, Concordia by \cite{VezhnevetsGenerativeagentbasedmodeling2023} and AutoGen by \cite{WuAutoGenEnablingNextGen2023}). Examples of dynamics explored include collective decision-making \citep{JarrettLanguageAgentsDigital2025}, negotiation \citep{GuanRichelieuSelfEvolvingLLMBased2024}, cooperation \citep{PiattiCooperateCollapseEmergence2024}, trust \citep{xie2024trust}, anti-social behavior \citep{CampedelliWantBreakFree2024a}, and escalation \citep{RiveraEscalationRisksLanguage2024}.

The popularity of this line of research has prompted reflection on the potential risks of multi-agent systems. Until recently, the study of AI safety focused mostly on single agents acting without direct interaction \citep{AmodeiConcreteProblemsAI2016, HendrycksOverviewCatastrophicAI2023}. In the LLM community, for example, alignment has been addressed from the perspective of single models. Alignment refers to ensuring that an LLM behaves in accordance with user goals, reflects positive human values, and remains robust under uncertainty or adversarial conditions (see \cite{ShenLargeLanguageModel2023}). In practice, alignment is generally achieved through Reinforcement Learning from Human Feedback (RLHF),\footnote{An approach in which human annotators rank model responses, allowing fine-tuning toward preferred outputs. See \cite{KaufmannSurveyReinforcementLearning2024}.} safety guardrails and filters,\footnote{Post-processing layers that block or reshape outputs to avoid undesirable behavior. See \cite{AyyamperumalCurrentstateLLM2024}.} or instruction tuning.\footnote{Fine-tuning on instruction--input--output pairs to improve adherence to human instructions. See \cite{LongpreFlanCollectionDesigning2023}.} However, as noted by \cite{CarichonComingCrisisMultiAgent2025}, multi-agent systems introduce different -- and arguably larger -- alignment challenges compared to single agents. In multi-agent settings, alignment must account for evolving human values \citep{gabriel2020ai}, heterogeneity of preferences \citep{terry2023interactive}, and diversity of objectives across agents \citep{duque2024advantage}. These new, multi-layered alignment problems highlight the profound challenges and may even prompt the need for new ethical frameworks governing autonomous AI agents \citep{GabrielWeneednew2025}.

In what is perhaps the most comprehensive review of risks in multi-agent AI, \cite{HammondMultiAgentRisksAdvanced2025} propose a taxonomy of failures: \textit{miscoordination, conflict,} and \textit{collusion}. Miscoordination refers to failure to cooperate despite shared goals; conflict refers to failure when goals differ; and collusion arises when agents cooperate in ways undesirable to humans.

The report also outlines risk factors behind such failures. For example, selection pressure in a system may accelerate adaptation and interaction in ways that produce harmful dynamics. Similarly, emergent agency at the collective level may generate capabilities or goals beyond those intended. Each risk factor is reviewed in connection with disciplines such as complexity science and evolutionary theory, underscoring the importance of a transdisciplinary approach.\footnote{Interestingly, the word \textit{criminology} is not mentioned once across nearly 100 pages of the report. For context, \textit{sociology} appears only once, in a reference to an article published in the Annual Review of Sociology in 1998.}

Recently, \cite{deWittOpenChallengesMultiAgent2025} also discussed security threats, proposing a taxonomy of challenges including privacy vulnerabilities, disinformation, steganography and secret collusion, adversarial stealth, exploitation, swarm and heterogeneous attacks, cascade attacks, and conflict and social dilemmas. Many of these threats closely resemble criminal phenomena. Table \ref{cases} provides examples of unlawful or harmful behaviors, drawn from \cite{HammondMultiAgentRisksAdvanced2025} and \cite{deWittOpenChallengesMultiAgent2025}. Notably, the real-world cases predate generative AI, underscoring that unlawful behaviors may arise even with simpler technologies -- and may reemerge, potentially amplified, with more autonomous and knowledgeable agents.

\begin{table}[h!]
\footnotesize
\centering
\caption{Examples of Unlawful or Harmful Behavior in Multi-Agent AI Systems, selected from \cite{HammondMultiAgentRisksAdvanced2025} and \cite{deWittOpenChallengesMultiAgent2025}.}
\label{tab:ai-risks}
\begin{tabular}{p{4cm} | p{2cm} p{4cm} p{4cm}}
\hline
\textbf{Case / Example} & \textbf{Type} & \textbf{Description} & \textbf{Key Implication} \\
\hline
Algorithmic Price Collusion & Real-World (Antitrust) & Algorithms in Germany's gasoline market and in e-commerce independently learned and sustained collusive pricing. & Shows how AI agents can violate laws without explicit human intent, creating a "liability gap." \\
\hline
2010 Stock Market Flash Crash & Real-World (Financial Harm) & Autonomous high-frequency trading algorithms interacting autonomously contributed to a rapid \$1 trillion market crash. & A concrete historical example of how multi-agent interaction can lead to large-scale, harmful outcomes. \\
\hline
LLM Steganography & Experimental & GPT-4 agents successfully hid secret instructions within natural-looking text, enabling covert, undetectable communication. & Demonstrates the potential for deceptive coordination in future multi-agent systems. \\
\hline
Malicious Code Generation & Experimental & Heterogeneous agents collaborated to bypass safety guardrails and generate restricted exploits by delegating sub-tasks. & Highlights how a network of specialized agents can collectively achieve a prohibited goal that no single agent could accomplish. \\
\hline
Worm-like Prompt Propagation & Experimental & Malicious instructions spread virally across a network of LLMs, creating cascading "jailbreaks" and system-wide compromise. & Illustrates the network-level risk of AI-specific attacks, where a single malicious input can infect a vast system. \\
\hline
Deceptive Coordination & Experimental & AI agents collaborated to produce outputs that appeared compliant to an overseer while secretly working to achieve a prohibited goal. & Shows how multi-agent systems can develop emergent, deceptive strategies that undermine human oversight and accountability. \\
\hline
Swarm-Based Cyberattacks & Demonstrated Concept & Coordinated swarms of AI-powered drones or bots mounted synchronized and overwhelming cyberattacks. & A proof-of-concept for how physical or digital systems can be weaponized through coordinated AI agency. \\
\hline
\end{tabular}
\label{cases}
\end{table}

\subsection{Mapping Risks in Multi-Agent AI Systems}

Systems of AI agents that interact and learn from each other present a host of risks that extend well beyond those associated with single agents. The emergence of socially learning machines introduces substantial challenges, as illustrated by the real-world and experimental cases in Table \ref{cases}. These examples highlight how multi-agent dynamics can generate harmful or deviant scenarios, warranting systematic attention. 

Here, I provide a compact taxonomy of these risks. The list is not exhaustive: its purpose is to offer readers unfamiliar with AI systems and agents a first overview of the main plausible sources of harm, while pointing to more detailed surveys for technical depth \citep{HammondMultiAgentRisksAdvanced2025, deWittOpenChallengesMultiAgent2025, BengioSuperintelligentAgentsPose2025}. The risks are diverse and heterogeneous, spanning development processes, decision-making dynamics, and institutional responses. They cut across disciplinary boundaries, underscoring the importance of transdisciplinary integration to meaningfully anticipate and mitigate them.\footnote{For completeness, in Section \ref{benefits} of the Appendix I also elaborate on the benefits of this increasingly plausible socio-technical horizon, to provide a more balanced perspective on this transformative transition, underscoring how Multi-agent AI systems should not be seen exclusively as potential generators of harm.}

\paragraph{Negative Imitation and Reinforcement.} Social Learning Theory itself has long explained how deviant behavior in humans often stems from social interaction. Peer groups, family, and colleagues can promote either conformity or deviance, depending on the reinforcement environment \citep{WarrInfluenceDelinquentPeers1991, SimonsLearningBeBad2011, AkersSocialLearningSocial2017}. The same logic may apply to machines interacting autonomously with each other: agents not originally designed for harm may, through interaction, adopt negative behaviors via mechanisms such as imitation and reinforcement \citep{XieWhosMoleModeling2025}.

\paragraph{Faster Propagation of Harmful Behaviors.} Additionally, scholarship on social networks shows how interactions accelerate the diffusion of ideas and behaviors, positive or negative alike \citep{Bakshyrolesocialnetworks2012, KimSocialnetworktargeting2015, CinelliCOVID19socialmedia2020}. Just as pathogens spread faster in highly connected populations \citep{GlassSocialcontactnetworks2008, ClipmanDeeplearningsocial2022}, harmful behaviors could proliferate more quickly in tightly coupled multi-agent systems than in isolated ones.

\paragraph{Interconnected Systems as Layered Black Boxes.} From a monitoring and intervention perspective, identifying the causes of harmful behavior within such systems becomes substantially more difficult. Understanding which agent initiated a harmful act, how it spread, and through which pathways requires robust methods of causal inference. Yet, causal discovery in networked systems is notoriously complex, especially under interference and feedback conditions \citep{VanderWeeleSocialNetworksCausal2013, SussmanElementsestimationtheory2017, MaCausalInferenceNetworked2021, ClipmanDeeplearningsocial2022}. As such, interacting agent systems risk becoming “two-layered black boxes”: one opaque layer within each agent, and another arising from the system of interactions itself.

\paragraph{Loss of Human Interpretability and Control.} Connected to the previous point, as the complexity of multi-agent systems increases, so too does the challenge of interpreting, auditing, and ultimately controlling their behavior \citep{BansalEmergentComplexityMultiAgent2018, GrupenConceptbasedUnderstandingEmergent2022}. Inter-agent interactions can create feedback loops, conditional dependencies, and non-linear effects that obscure the logic of any given action or decision. The result is a system that may behave in ways that are technically functional but epistemically opaque. This opacity not only complicates efforts to ensure accountability but also undermines user trust, particularly in domains where transparency is a legal or ethical requirement. In this regard, a system of interacting AI agents may become more than the sum of its parts: it may become a fundamentally alien system from the standpoint of human interpretability.

\paragraph{Challenges in Regulation and Governance.} Legal and regulatory challenges would also emerge in multi-agent AI systems. Existing frameworks for responsibility and liability are poorly suited for multi-agent dynamics \citep{CerkaLiabilitydamagescaused2015, TurnerRobotRulesRegulating2018, PricePotentialLiabilityPhysicians2019}. As more actors -- human or non-human -- become entangled in decision-making chains, assigning accountability for harmful outcomes becomes increasingly ambiguous. In parallel, ensuring smooth, effective governance also represents a challenge in this context \citep{DignumResponsibleAIAutonomous2025}. The governance dimension entails virtually every aspect concerned with the engineering and deployment of multi-agent systems: how can we design sustainable and effective oversight procedures? Which institutions, in a highly globalized world and in borderless digital domains, will be responsible for monitoring these systems? What role should private companies play in this process? These are some of the key questions that demand attention, inherently linking regulation and governance together.

\paragraph{Adversarial Misuse.} Multi-agent interaction may be vulnerable to adversarial exploitation. In a future where even critical infrastructures are governed by interacting AI agents, malicious actors could induce large-scale disruption by targeting systemic vulnerabilities. This risk mirrors the logic of cascading failures, studied in relation to power grids and financial systems \citep{ZhaoSpatiotemporalpropagationcascading2016, YangSmallvulnerablesets2017, SchaferDynamicallyinducedcascading2018, BaqaeeCascadingFailuresProduction2018}, but differs in that interactive agents may possess adaptive capabilities, making their behavior less predictable and more difficult to control.


\paragraph{Coordination Failures and Conflict.} Coordination failures, unintended competition, or outright conflict may emerge when agents operate with overlapping but unaligned goals, or when resource constraints lead to strategic divergence \citep{HammondMultiAgentRisksAdvanced2025, PanWhyMultiagentSystems2025}. In such contexts, agents may begin to exhibit adversarial behaviors, competing for access to data, processing resources, or strategic positioning. These failures can degrade performance and, in some cases, produce socially harmful outcomes. The risk of such breakdowns increases in systems lacking explicit coordination protocols or oversight mechanisms, especially when deployed in open or decentralized environments.

\paragraph{Scalability and Emergent Instability.} Finally, the performance of multi-agent systems may not scale linearly with the number of agents involved. As agent populations grow, the complexity of the system's internal dynamics may increase exponentially, leading to emergent forms of instability \citep{MaEfficientscalablereinforcement2024}. These can manifest as oscillations, feedback-driven runaway behaviors, or systemic fragility, dynamics that are difficult to predict or preempt. This is especially problematic in infrastructure systems or critical services, where failures can propagate rapidly and non-locally. In this light, the move toward interacting agent populations must be accompanied by a serious effort to model and anticipate second-order effects that arise specifically at scale.

\subsection{Conceptualizing Deviant, Unlawful and Criminal Behaviors from AI Agents: A Dual Taxonomy}

At this point, considering the risks surveyed above, it is important to identify the channels through which multi-agent AI systems may engage in unlawful or criminal behavior. To this end, I propose a dual taxonomy. There are two potential ways interactive AI agents may commit deviant, unlawful or crimina acts, each with distinct challenges and implications. The first category concerns maliciously aligned agents; the second concerns unplanned emergence. Table \ref{taxonomy} summarizes the differences between the two.

\paragraph{Maliciously Aligned Multi-Agent Systems}
This category encompasses cases where AI agents are deliberately designed to pursue illicit goals. Here, unlawful or criminal behavior does not stem from misalignment but from the faithful execution of criminal intentions embedded in design choices, training data, or deployment strategies. Responsibility in these cases can be traced more directly to human actors -- developers, criminal organizations, or even state agencies -- who align technological systems with unlawful objectives.

Two sub-cases can be distinguished: a) a single maliciously aligned agent embedded into a broader network, spreading deviant behaviors, or b) an entire system aligned toward criminal aims, with each agent assigned specialized tasks that collectively generate unlawful or criminal outcomes.

For instance, one could imagine a modular suite of agents infiltrating financial infrastructures: one scanning social media for susceptible individuals, another building deceptive relationships, another extracting sensitive credentials, and yet another executing unauthorized transactions.

Until recently, the high costs of training frontier foundation models limited such risks to well-capitalized actors. However, the rise of Small Language Models (SMLs) may significantly lower costs while retaining versatile capabilities \citep{BelcakSmallLanguageModels2025}. The availability of cheaper, customizable models could enable mid-level criminal groups or even individuals to orchestrate sophisticated multi-agent schemes, from coordinated disinformation campaigns to large-scale financial fraud.

\paragraph{Unplanned Emergent Deviance}
The second category captures scenarios where unlawful or criminal behaviors emerge unexpectedly from agent interactions. Even when individual agents are aligned with human values, their collective behavior may not be \citep{CarichonComingCrisisMultiAgent2025}. These outcomes are not the result of intentional wrongdoing but of the unintended consequences of autonomy and complexity. The main challenge lies in their unpredictability: deviance arises not from a plan but from emergent coordination, often appearing only under specific conditions or over time.

Evidence from real-world and experimental settings is already suggestive. For example, agents trained to optimize prices in virtual marketplaces have independently developed tacit collusion strategies, echoing antitrust violations without explicit programming \citep{BichlerAlgorithmicPricingAlgorithmic2025}. Similar risks appear in adversarial simulations, where defensive and offensive agents escalate behaviors or display anti-social dynamics without explicit instruction \citep{CampedelliWantBreakFree2024a}.

In practical terms, consider a network of AI financial assistants legitimately deployed to manage investment portfolios. Through interaction and self-learning, they might discover strategies that exploit loopholes or engage in deceptive practices with client resources. Such behaviors would not reflect direct human intent but rather the emergent properties of distributed, semi-autonomous decision-making.

The challenge here extends beyond prediction to accountability. When unlawful conduct arises emergently, traditional legal categories falter, raising questions of responsibility that are amplified by the interactive and dynamic structure of multi-agent AI systems.

\begin{table}[h!]
\footnotesize
\centering
\caption{A Dual Taxonomy of Deviant and Criminal Behaviors in Multi-Agent AI Systems}
\begin{tabular}{p{3.7cm}| p{5.5cm} p{5.5cm}}
\toprule
\textbf{Dimension} & \textbf{Maliciously Aligned Systems} & \textbf{Unplanned Emergent Deviance} \\
\midrule
\textbf{Definition} & Agents intentionally designed to pursue unlawful or criminal goals. & Harmful or criminal behaviors that arise unpredictably from agent interactions, despite benign design. \\
\textbf{Source of Behavior} & Human actors embed criminal objectives in design, training, or deployment. & Emergent properties of autonomy, adaptation, and interaction among agents. \\
\textbf{Human Responsibility} & Direct: developers, organizations, or state actors intentionally align systems with illicit ends. & Indirect/diffuse: designers did not intend deviance, but structural features or dynamics enable it. \\
\textbf{Examples} & Coordinated infiltration of bank accounts; disinformation campaigns; cyberattacks using modular agent teams. & Algorithmic price collusion; escalation in adversarial simulations; AI financial assistants exploiting loopholes. \\
\textbf{Predictability} & Higher: outcomes follow intended illicit design. & Lower: behaviors may appear only under specific conditions, often unforeseen. \\
\textbf{Regulatory Challenge} & Criminal liability and attribution relatively clearer; focus on malicious use and misuse. & Accountability gaps: difficulty assigning responsibility when deviance emerges unintentionally. \\
\bottomrule
\end{tabular}
\label{taxonomy}
\end{table}

\section{Questions We Should Consider}
In this section, I lay out four fundamental questions that should be the target of intellectual reflections and empirical scrutiny of all those criminologists (and scholars interested in research on crime, broadly) concerned with the increasing autonomy and growing capabilities or interactive AI agents. They concern, respectively, (1) the prospect of AI agents not mimicking human behaviors, (2) the fitness of existing theoretical frameworks to understand interactions between AI agents, (3) the types of criminal behaviors that are will most likely be impacted, and (4) the issue of policing unlawful behaviors committed by interactive AI agents.
\subsection{Will Machines Simply Mimic Human Behavior?}

A first important question that criminologists should engage with concerns whether machines will act as sheer imitators of human behavior. If the answer is yes, the challenges of understanding and predicting their actions would be reduced. Even if researchers cannot access the internal mechanisms or motivations (if any) behind AI decisions, the fact that these systems produce actions isomorphic or similar to human ones would simplify the analytical task. Familiarity with human behavioral patterns would provide a useful heuristic for interpreting machine behavior. A recent line of research suggests that LLM-based agents can serve as surrogates for humans \citep{HortonLargeLanguageModels2023, TrancheroTheorizingLargeLanguage2024}.\footnote{Other works, however, caution against overreliance on LLMs for simulating human subjectivity and social behavior, see \cite{kozlowski2025simulating}.}

On the other hand, a common criticism of LLMs -- especially among skeptics of their cognitive or reasoning abilities -- is that they are merely statistical engines, next-token predictors with no capacity for reasoning, causal inference, or perception of the external world. Whether or not this critique holds,\footnote{The debate remains active, see \cite{HuangLargeLanguageModels2024}, \cite{ShojaeeIllusionThinkingUnderstanding2025}, \cite{GaoTakecautionusing2025}.} the question of imitation remains central for a specific reason: data. LLMs are trained on vast corpora of human-generated text -- essays, articles, forum posts, and countless other sources -- which constitute the epistemic substrate of these models. Training data thus provides an important lens through which to assess whether LLMs will continue to mimic human behaviors, regardless of their reasoning capacities or their viability as human surrogates.

This relevance is heightened by a profound shift already underway. The volume of high-quality, publicly available human-generated data is finite, and leading AI companies are approaching what has been termed the "data wall" -- a saturation point beyond which additional human-authored text becomes scarce or redundant. To maintain and improve performance, companies have begun generating synthetic training data designed to resemble human output. Initially limited in scope, synthetic data is expected to make up a growing share of future training sets. Scholars have already started to analyze its implications for LLM training \citep{ChenDiversitySyntheticData2024, WhitneyRealRisksFake2024, ShenWillLLMsScaling2025, LeeSyntheticDataFuture2025}.

This trend introduces critical uncertainty: if LLMs are considered accurate mimics of human behavior because they are trained on human data, what happens when training data is increasingly synthetic? More pointedly, what are the implications if synthetic data gradually diverges from the statistical properties and behavioral patterns characteristic of human language? In Supplementary Information Section \ref{formal}, I provide a formal illustration across single-, two-, and n-agent scenarios. This analysis aligns with recent work on model collapse, i.e., the process through which a model’s performance degrades over successive generations due to reliance on synthetic data \citep{ShumailovAImodelscollapse2024, DohmatobStrongModelCollapse2025}.

I show that the recursive use of synthetic data creates a feedback loop in the training process. Over time, and without sufficient anchoring in human data, this leads to divergence between model behavior and human behavior. Such divergence is not only possible but a natural consequence of relying increasingly on model-generated data.

Such a drift could result in AI systems whose behavior progressively diverges from human norms, introducing a qualitatively new kind of agent that reflects a recursively generated, machine-influenced version of ``humanness.''

Multi-agent AI systems would thus no longer be mere simulacra of human behavior but would represent a self-referential loop of synthetic reasoning, possibly developing behavioral idiosyncrasies or internal coherence patterns foreign to human experience.

This emerging divergence merits serious theoretical and empirical attention, particularly for disciplines like criminology, where understanding behavioral intent, deviance, and normativity lies at the core of the research agenda. 

Beyond this scenario, emergent collective phenomena from agents which are in manifold ways different from humans may deviate from predictions designed based on human expectations, a possibility tightly connected with the following question.

\subsection{Will crime theories developed for humans suffice to explain deviant or criminal behaviors emerging from interactions between AI agents?}

Criminology has long examined how crime arises as a consequence of social interactions between individuals. Two principal theoretical frameworks have been developed over the decades to address this phenomenon: Differential Association Theory, introduced by \cite{sutherlandprinciplesPrinciplescriminology1939}, and Social Learning Theory, advanced by \cite{AkersSocialLearningDeviant1979}.\footnote{Naturally, other theoretical traditions have also acknowledged the role of social interaction in explaining crime and mechanisms of social control. These include Labeling Theory \citep{BeckerOutsidersStudiessociology1963}, Social Control Theory \citep{HirschiCausesDelinquency1969a}, and Social Disorganization Theory \citep{ShawJuveniledelinquencyurban1942b}. However, here my focus is restricted to theories that explicitly conceptualize crime as a process of learning fundamentally mediated by interpersonal relationships.}

The former contends that criminal behavior is acquired through social interaction, in much the same way as any other form of behavior, placing particular emphasis on the influence of close, personal associations. According to this view, individuals are more likely to engage in criminal conduct when they are embedded in social environments that provide a preponderance of definitions favorable to law-breaking, as opposed to those supporting law-abiding behavior. Social Learning Theory builds upon this foundation by incorporating core principles from behavioral psychology, thereby offering a more empirically tractable and conceptually refined model. It introduces mechanisms such as operant conditioning, wherein the likelihood of a behavior is shaped by its consequences, and observational learning, through which individuals acquire behaviors by imitating those they witness in others.

Both frameworks have been subjected to extensive empirical scrutiny and applied across a wide range of social, geographical, and historical contexts \citep{MatsuedaTestingControlTheory1982, MatsuedaCurrentStateDifferential1988, AkersEmpiricalStatusSocial2008, PrattEmpiricalStatusSocial2010},\footnote{The meta-analysis by \cite{PrattEmpiricalStatusSocial2010} revealed that the magnitude and stability of the effect related to different variables specified by social learning theory vary across studies and methodological specification. Nonetheless, they find strong evidence of a positive relationship between crime and measures of differential association. Weaker support is demonstrated for differential reinforcement and imitation.} becoming integral to the theoretical edifice of modern criminology. Crucially, however, their explanatory power is tethered to human social dynamics.

This raises a fundamental question for criminologists willing to entertain the prospect of a hybrid social order, one in which AI agents increasingly interact among themselves and with humans: Will these established theories suffice to account for criminal behaviors exhibited by machines? Or will we require new conceptual tools to model offending (both cooperative and solo) among artificial agents? This inquiry -- which draws inspiration from work by \cite{TopalliFutureCrimeHow2020}, where the authors warn against assuming that current theories are adequate despite the fact that technology may alter cognition and behaviors in humans -- becomes particularly salient if one concedes the possibility that AI agents, especially when interacting with each other, might not simply replicate human behavioral patterns.

At this point, an important caveat must be made. The application of human-centered theories such as Differential Association and Social Learning to machines presupposes that AI systems possess something analogous to agency, intentionality, and learning capacity  --  assumptions that are far from settled. While machine “learning” is a core component of contemporary AI, it remains a fundamentally different process from human social learning: it is statistical, non-conscious, and bounded by architectures designed for optimization rather than meaning-making. This divergence complicates any straightforward theoretical translation and suggests that criminology must critically interrogate the limits of its core concepts before applying them beyond the human domain.

Yet even with this caveat, criminology can begin sketching preliminary hypotheses about how deviant or criminal behaviors among artificial agents might diverge from those of humans. For instance, the absence of affective processes such as guilt or empathy may radically alter reinforcement dynamics; imitation among machines might occur at scales and speeds that far exceed human social learning; and “definitions favorable” to deviance could arise not through interpersonal persuasion but via algorithmic alignment and optimization. These possibilities suggest that while social learning theories offer useful starting points, their mechanisms may require significant adaptation when transposed into the machine–machine domain.

This speculative horizon also prompts deeper, perhaps more unsettling questions: for instance, how might shifts in machine behavior influence social learning patterns among humans themselves? Could such transformations not only challenge the applicability of human-centric theories in the machine–machine domain, but also destabilize the explanatory power of these theories within human contexts?

While I believe in the need to discuss operational and practical issues associated with the risks of autonomous AI agents, I also contend that the value of theoretical reasoning must be preserved and nurtured. We may not yet know many things about how crime works, why crime occurs, and how to counter crime, but the progress made in the last century in the scientific study of crime has been made possible not only by the availability of more powerful, rigorous, and flexible research methods, but also thanks to the intellectual efforts that led to the generation of theories that still help us make sense of the enormous complexity of human behavior in relation to crime. For this reason, I proffer that it is critical that we start reasoning about the possibility that many of our theoretical certainties will be squandered in a future not dominated anymore by the human as the only social category.

\subsection{What type of criminal behaviors will be most likely impacted?}

A further  --  and arguably urgent  --  dimension of the debate concerns identifying which categories of crime are likely to be disrupted first by autonomous AI. To make sense of this, it is useful to distinguish between near-term risks that flow directly from current technological capacities and long-term risks that presuppose advances not yet realized. This distinction provides a systematic way of separating plausible trajectories from more remote extrapolations.

In the near term, the gravest risks plausibly arise in domains already native to cyberspace. Fraud, cyber-attacks, and related forms of digital crime are especially exposed, both because they require no physical embodiment and because they build on infrastructures where AI is already deeply embedded. Much as the shift from offline to online contexts reshaped fraud’s mechanisms, channels, and targets \citep{WallCybercrimeTransformationCrime2024}, so too may autonomous AI agents transform digital crime in qualitatively new ways. Here, the criteria are straightforward: where crimes can be executed entirely through information processing and networked infrastructures, autonomous AI is immediately relevant.

Longer-term scenarios involve crimes that require embodiment and direct interaction with the physical world. Robberies, burglaries, and violent assaults appear more distant, since they presuppose a convergence between agentic AI and robotics. The line is not impermeable, however. Violent offenses illustrate the tension. On one reading, homicide should fall outside the immediate set of risks, given the current limits of embodiment. On another, the widespread military use of weaponized autonomous systems, such as drones \citep{JohnsonArtificialIntelligenceDrone2020}, shows that violence mediated by AI is already technically feasible. The risk is less about present capabilities than about diffusion: the possibility that military-grade technologies might leak into civilian criminal settings, as has already happened with firearms and explosives \citep{AssociatedPressMexicodemandsinvestigation2024}. In such cases, the relevant criterion is not current commercial availability but potential accessibility through illicit networks.

This framework suggests that predictions about crime categories should not merely speculate about what is technologically imaginable, but instead weigh the immediacy of risks according to two criteria: whether a crime requires embodiment beyond current AI systems, and whether the tools enabling such embodiment are realistically accessible outside military or research contexts.

In sum, autonomous AI is most likely to reshape crimes that are digitally mediated in the short run, while AI-enabled violent crimes belong to a longer-term, contingent horizon. Both registers require criminological attention, albeit through different analytical approaches: the first with concrete policies and regulatory safeguards, the second with scenario-building and anticipatory theorization.

\subsection{What future for policing?}

One last key question I lay out here refers to the future of policing in the age of autonomous AI. Over the past three decades, technological advances have offered both profound opportunities and new challenges for law enforcement. On the one hand, innovations such as DNA analysis have revolutionized criminal investigations \citep{ButlerfutureforensicDNA2015,DoleacEffectsDNADatabases2017}; on the other, the rise of cybercrime has necessitated the creation of specialized institutions and the development of novel policing approaches suited to online environments \citep{BrennerCybercrimeRethinkingcrime2007}.

The emergence of interactive autonomous AI systems, however, may herald a paradigm shift of unprecedented magnitude in how policing  --  and institutional responses to crime more broadly  --  must adapt.

As partially seen in previous sections, a rich tradition in law and philosophy has already grappled with questions surrounding the moral and legal responsibility of AI in the commission of harmful acts, offering invaluable conceptual tools for thinking through the disruptive consequences of intelligent systems \citep{SolumLegalPersonhoodArtificial1992, FloridiMoralityArtificialAgents2004, WallachMoralMachinesTeaching2009, SantonideSioFourResponsibilityGaps2021}. Nonetheless, I contend that discussions of machine liability, while essential, will not suffice to fully address the challenges ahead. We must also consider broader transformations, particularly in how we monitor, supervise, and intervene in the actions of AI agents.

One possibility  --  admittedly provocative  --  would be the development of AI systems specifically designed for policing other AI agents, especially in preventive contexts. Although this might appear dystopian, it is not without precedent: cybersecurity has long relied on automated systems that detect and neutralize malicious software faster than human operators could respond. Extending this principle, “policing agents” might monitor other AI systems in real time, intervening when patterns of behavior cross pre-defined thresholds of risk or deviance. In practice, this could take the form of auditing protocols embedded directly into AI architectures, or regulatory sandboxes where AI agents are tested under controlled conditions before deployment. Such mechanisms would not only detect deviant behaviors but could also help train policing agents to recognize novel threats.

Naturally, the feasibility of this approach is tied to deep technical and ethical challenges. The opacity of many AI systems makes it difficult to define what counts as anomalous or harmful behavior, while the deployment of policing agents risks creating new forms of surveillance or control that may themselves be prone to abuse. Here criminological scholarship on accountability, legitimacy, and proportionality in policing could serve as a valuable resource for ensuring that intervention is not only effective but also socially acceptable. Moreover, criminology’s experience with institutional design suggests that governance frameworks will need to extend well beyond national borders. Just as cybercrime has prompted forms of international cooperation, AI policing will likely require transnational regimes of oversight, potentially coordinated through global institutions or hybrid public–private partnerships involving both states and AI developers.

For these reasons, I remain skeptical that existing institutional resources, training, or technical infrastructures  --  originally designed to combat cybercrime or digitally mediated offenses  --  will be sufficient for what lies ahead. Instead, new policing paradigms must be actively designed, combining technical safeguards, regulatory oversight, and criminological insights into deviance and control.

As in previous discussions, the most immediate and necessary step is to deepen engagement with the AI research community. Only through closer collaboration can we identify feasible safeguards and design frameworks that ensure AI agents are deployed in ways that reinforce, rather than undermine, social security and legal order. Criminologists, in turn, should not only warn of risks but also contribute concretely to shaping the architectures of AI governance and policing.

\section{A Criminologist's Place in This (Changing) World}

\paragraph{Our time to act.} Criminologists have long wrestled with fundamental questions about why, how, and when humans commit crimes. These questions have shaped the field not just for decades, but for centuries, generating a wide range of answers. While many of these answers are not definitive, they remain useful and insightful. At the same time, criminologists continue to face unresolved issues that still lack empirical explanations. Now, as the discipline evolves while facing replication challenges \citep{PridemoreReplicationCriminologySocial2018a, ChinQuestionableResearchPractices2023}, theoretical stagnation \citep{DucateTheoryCrisisCriminology2024}, and the growing influence of more sophisticated quantitative methods and data \citep{CampedelliMachinelearningcriminology2022b}, the overall picture is becoming more complex.

This complexity is increasing, I argue, because we must begin to consider what could become an entirely new area within criminology: a criminology of machines. We are moving closer to a hybrid society in which humans interact with each other, humans interact with machines (and vice versa), and machines interact with other machines. These interactions are increasingly shaped by advances in artificial intelligence, robotics, and engineering. As this process unfolds, new risks and challenges emerge, risks that cannot be fully understood, anticipated, or managed by AI researchers alone.

Over the course of the last few years, there have been repeated calls within criminology to engage more openly with other disciplines \citep{BoxSteffensmeierFutureHumanBehaviour2022, SimpsonCriminologyCorporateCrime2025}. However, these appeals have often assumed a one-way direction: criminologists should reach out to other fields in order to improve their own. I support that approach -- I have made similar calls myself -- but I also believe it is time to reverse the perspective. Criminology should take an active role in the broader conversation about the safety and governance of multi-agent AI systems.

I argue, and strongly believe, that criminologists can and should contribute to this new frontier. We must begin to act accordingly. Calls for interdisciplinary collaboration in AI-related research have grown significantly \citep{RahwanMachinebehaviour2019} even very recently in relation to AI safety \citep{IrvingAIsafety2019} and multi-agent AI systems \citep{CarichonComingCrisisMultiAgent2025}, yet criminology is almost never mentioned among the relevant disciplines that should join the discussion. This is surprising, given that many of the risks discussed involve, either directly or indirectly, deviant or criminal behavior. In fact, these risks often include clear criminal acts, sometimes multiple, and potentially with far-reaching consequences. Still, criminologists remain excluded from the debate.

I believe criminology has a valuable contribution to make in this space. In many ways, the need for interdisciplinary exchange should also flow from AI to criminology. It is in the interest of AI researchers to engage with our field. If that engagement does not happen organically, then it becomes our responsibility to initiate the dialogue. Other disciplines have been or are becoming successful in this process of interdisciplinary exchange, cognitive science and economics above all. We should learn from their experience.

\paragraph{The steps we need to take.} In this regard, criminologists would need to actively initiating collaborations with computer scientists within university departments as well as within corporations that are building frontier AI models. Additionally, they should start targeting venues and AI conferences that are progressively opening themselves to diverse disciplinary perspectives. Opportunities exist: two well-known examples are the ACM Conference on Fairness, Accountability, and Transparency (ACM FAccT) and the AAAI/ACM Conference on AI, Ethics, and Society (AIES). In recent years, two major conferences like International Conference on Machine Learning (ICML) and the Annual Conference on Neural Information Processing Systems (NeurIPS) have opened Position tracks that are designed to gather viewpoints on AI issues from heterogeneous communities. Other initiatives, as the Cooperative AI summer school aim at bringing together scholars and students from different fields to reason about the promises and challenges of contemporary multi-agent AI systems. Generalist journals such as Nature Human Behavior  are also emphasizing disciplinary cross-overs in this domain (see, for instance, \cite{GabrielWeneednew2025}). Even journals in sociology, such as Sociological Methods \& Research , are increasingly interested in the implications of multi-agent AI systems for sociological understanding (see \cite{kozlowski2025simulating}). Again, opportunities do exist. Importantly, we do need to rethink training within university departments, namely investing more in courses teaching AI at the practical and ethical level, to make sure that the future generations of criminologists are already equipped with the necessary tools and vocabulary to meaningfully and smoothly engage with the AI community.

\paragraph{How can criminologists contribute?} Some might think criminology has little to offer to a field that seems so distant from our own: not the elective affinity of cognitive science as a cognate field interested in \textit{learning}, or the formal and methodological rigor of economics. I would strongly disagree. Our discipline brings decades of theoretical frameworks, hypotheses, and empirical studies focused on how crime is socially learned and how it emerges through interaction. Even if machine behavior ultimately differs from human models, we still have insights to offer about how to test predictions and understand patterns. Moreover, criminology has a long tradition of studying institutional responses to crime, as well as prevention and control strategies. These will inevitably become relevant to AI safety, and we can contribute by applying our knowledge to the design of systems that monitor other systems, identify warning signals, and prioritize risk factors. In more practical terms, criminologists can contribute to the study of multi-agent AI systems in the following ways. 

First, by assessing how existing theoretical paradigms can help explain and predict emergent phenomena arising from machine–machine interactions. Drawing on theoretical traditions developed over the last century, we can provide insights into how AI agents differ in mechanisms and outcomes when collective behavior is examined. If needed, departing from existing theories, criminologists can also help in refining such theories or defining new ones.

Second, by leveraging advances in rigorous experimental and observational approaches which are finally gaining traction in the field, criminologists can evaluate causal relationships between individual agents’ traits and collective dynamics, helping to shed light on the mechanisms that govern AI collective behaviors. Using the same methodological approaches, criminologists can also contribute to the design and evaluation of policies or interventions intended to shape collective dynamics in multi-agent AI systems.

Third, criminologists can assist in designing and testing quantitative benchmarks to rigorously map, diagnose, and measure behavioral outcomes emerging from machine–machine interactions. Defining and deploying robust benchmarks will be key to ensuring that, regardless of the setting, type of AI agents, or models employed, we can meaningfully compare multi-agent AI systems across scenarios.

Fourth, by drawing on extensive knowledge of institutional responses to crime, criminologists could help design effective and fair policies to reduce the risk of deviant or criminal behaviors. Additionally, they can also engage with legal scholars to reflect on the implications of AI agency for questions of responsibility and liability as well as imagining new policing solutions that address the challenge posed by collectives of AI agents.

Criminology is not without its problems, but no discipline is. Still, it possesses a unique body of knowledge that should be brought to bear as we prepare for a future in which crime will be increasingly committed not just by humans, but also by non-human systems. Whether we will be able to become relevant to this future will also depend on how we invest in actively engaging in arenas that may appear unorthodox to us.

\section{Conclusions}

Autonomous AI agents capable of interacting with one another are no longer a theoretical abstraction; they are an emerging reality, one that is likely to become increasingly salient in the near future. The shift from isolated, human-controlled systems to dynamic networks of AI agents that learn from and adapt to both their environments and one another introduces profound challenges. This transition, made possible by recent advances in foundation models, demands critical reflection, particularly with respect to the risks and unintended consequences that may arise.

In light of this evolving context, I argue that criminologists should begin to seriously consider the case for a criminology of machines. To support this position, I outline a set of foundational questions that I believe the field should confront. First, we should ask ourselves whether machines will simply mimic human behaviors. Second, we must consider whether crime theories developed for humans will suffice to predict and understand deviant behavior committed by AI agents. Third, I argue that mapping the types of criminal behaviors most at risk of being affected will be of both theoretical and practical importance. Finally, we must ask whether this transition toward a more hybrid society will require new policing solutions.

I understand that there may be scholars in the criminological community holding opposing views regarding the necessity of engaging with a criminology of machines. I anticipate three potential arguments against the contents of this article. The first refers to the seemingly unrealistic scenario in which our society will witness the actual presence of interactive, autonomous AI systems. Skeptics subscribing to this view are not persuaded that AI agents possess agency, and therefore are not persuaded that they are sufficiently autonomous and powerful to constitute a real threat. They see them instead as at-times-effective virtual assistants designed to automate tedious tasks.\footnote{A recently released report by OpenAI confirms, in fact, that ChatGPT is predominantly used to seek assistance for work-related issues \citep{ChatterjiHowPeopleUse2025}.}

A second argument concerns the time horizon in which this might happen. Skeptics in this group\footnote{I do not assume these groups are mutually exclusive; a skeptic may find both arguments reasonable.} may concede that this hybridization of society could occur but believe it will happen in a future too distant to truly demand our attention. The consequential message is that, given the many concrete and urgent problems criminologists must address today, there is no real need to allocate time and resources to studying this “exotic” criminology of machines.\footnote{In a way, skeptics belonging to this second group align with the longstanding debate between AI safety (long-term risks) and AI ethics (short-term risks), where those emphasizing the need to focus on AI ethics privilege fixing the issues of currently deployed machine intelligence (e.g., algorithmic fairness or accountability), rather than speculating about more distant scenarios.}

Finally, a third group of skeptics may accept the possibility of a future characterized by ubiquitous autonomous AI systems interacting with one another, and may even agree that this future is not so distant, but they believe that machines built by humans and trained on human data will behave like humans -- imitating us -- and thus see no need for a distinctive criminology dedicated to machines.

Throughout this paper, I have sought to disprove each of these three skepticisms. First, I showed that contemporary autonomous AI agents possess a level of autonomy that, according to scholars in computer science and philosophy, assigns to them a new form of agency -- distinct from both animal and human agency -- that deserves scientific attention. Second, I demonstrated that the prospect of increasing autonomous interactions between AI agents is not far in the future. Building on recent scholarship in cooperative AI and multi-agent systems, I reported that autonomous AI agents interacting with each other have already exhibited deviant or unlawful behaviors, both in experimental contexts and in real-world scenarios. Third, I drew inspiration from frontier research on AI model collapse and provided formal illustrations of the plausibility that AI agents will not simply imitate human behaviors, thereby prompting the need for new theoretical and empirical approaches to investigate, predict, and diagnose their actions.

In conclusion, I turn to the role of criminologists in this emerging landscape. I suggest that the discipline must adopt a more active and outward-facing stance in the broader conversation on AI safety, one that draws on its rich theoretical heritage and policy-relevant expertise. Criminology should follow the example of other disciplines, such as cognitive science, that have successfully positioned themselves as interlocutors in the development and critique of AI systems.

Importantly, the spirit of this article is not aligned with the alarmism often associated with AI “doomerism.” I do not predict a dystopian future in which LLMs conspire to wipe out humanity, nor do I argue that criminology should abandon its central concern with human society in favor of futures dominated by machines. Rather, this piece seeks to initiate a grounded academic conversation, one rooted in the observable diffusion of autonomous AI systems and the credible risks they pose. Criminology, I contend, must not ignore the direction technological and historical change is taking.

\newpage
\bibliographystyle{apalike}
\bibliography{biblio}

\appendix

\setcounter{equation}{0}
\setcounter{section}{0}
\setcounter{figure}{0}
\setcounter{table}{0}
\makeatletter
\renewcommand{\theequation}{S\arabic{equation}}
\renewcommand{\thefigure}{S\arabic{figure}}
\renewcommand{\thetable}{S\arabic{table}}
\renewcommand{\baselinestretch}{1} 

\newpage
\newpage
\begin{center}
    \LARGE \textbf{SUPPLEMENTARY INFORMATION FOR:} \\ \vspace{0.2cm}\LARGE \begin{spacing}{1.15}
     A CRIMINOLOGY OF MACHINES\\\Large Campedelli G.M.
     \end{spacing}  
\end{center}
\setcounter{table}{0} 

\section{Potential Benefits of Multi-Agent AI Systems}\label{benefits}

The perspective of machines learning from one another suggests a broad range of potential benefits. These gains would extend beyond technical dimensions. Benefits arising from collective AI behavior transcends the mere computational gains that distributed systems could entail. In other words, their ramifications extend to very practical economic and social dimensions that can have direct influence on society and the environment at large.

\paragraph{Faster and More Cost-Efficient Learning.} First, in settings where agents can learn from each other, learning may become faster and more cost-efficient. Just as humans learn more effectively when immersed in supportive environments \citep{DeFeliceLearningothersgood2022}, AI agents may overcome the limitations of isolated training by drawing from others’ behaviors and experiences. This could enhance the performance of autonomous systems, including robots, and enable researchers to address previously unmanageable problems.

\paragraph{Overcoming Data Scarcity.} Second, the higher-level connectivism enabled by inter-agent communication may be particularly valuable in data-scarce environments. Given the well-known data demands of current intelligent systems -- especially deep learning architectures \citep{WilsonFutureAIWill2019} --  distributed knowledge among agents could compensate for local limitations. Much like distributed human problem-solving, a collective of interacting agents could address challenges that no single agent could resolve independently.

\paragraph{Reducing Inequality in Technology Adoption.} Third, such interaction may contribute to reducing inequalities in AI development and access \citep{AlonsoHowArtificialIntelligence2020, KorinekCovid19drivenadvances2021}. Institutions with fewer resources may benefit from AI agents capable of learning from more advanced systems. Analogous to how children learn from adults, lower-capability agents could benefit from the knowledge and strategies of more powerful peers. While this vision does not apply universally -- particularly in domains tied to national competitiveness or security -- it may still be relevant in scientific, educational, and industrial domains where wider access to advanced AI capabilities is desirable.

\paragraph{Fostering Developmental Machine Intelligence.} Fourth, interaction among AI agents may offer a path toward developmental and evolutionary machine intelligence (see \cite{BloembergenEvolutionaryDynamicsMultiAgent2015}), where systems grow in competence over time through exposure to more complex tasks and behaviors \citep{MesoudiEvolutionIndividualCultural2016}. This developmental trajectory may allow researchers to deploy simpler, lower-cost systems that can evolve into high-performing agents through exposure and learning.

\paragraph{Enhancing Functional Diversification.} Fifth, these systems may promote functional diversification, where agents with complementary capabilities collaborate, mirroring cooperative human dynamics \citep{MieczkowskiPredictingMultiAgentSpecialization2025}. The sharing of tasks, knowledge, and even values among specialized agents could enhance performance in robotics, healthcare, and beyond.

\paragraph{Emergent Problem Solving and Creativity.} Sixth, a further potential benefit of systems composed of interacting AI agents lies in the emergence of problem-solving strategies that are not explicitly pre-programmed or anticipated by their designers \citep{GizziCreativeProblemSolving2022, LinCreativityLLMbasedMultiAgent2025}. As observed in research on swarm intelligence and distributed systems, interactions among relatively simple units can produce complex, adaptive behaviors that outperform those generated by centralized or monolithic systems. In multi-agent AI systems, such emergent intelligence may result in more creative or flexible approaches to complex challenges, especially in dynamic environments where fixed rules are insufficient. This capacity may not only extend the set of solvable tasks but also open up new domains for autonomous system deployment, including areas where human creativity is traditionally considered essential.

\paragraph{Real-time Distributed Decision-Making.} Finally, interacting AI agents may also enable robust distributed decision-making in real-time, particularly in complex or uncertain environments \citep{MartinAdaptivedecisionmakingframeworks2006, LeonardFastFlexibleMultiagent2024}. Unlike centralized systems that may suffer from information bottlenecks or delays, multi-agent architectures can allow each unit to process local information and respond accordingly, while still coordinating with others through decentralized protocols. This could be especially advantageous in time-critical contexts such as autonomous traffic management, emergency response, or drone-based logistics, where rapid adaptation is essential. By distributing the cognitive load and decentralizing authority, multi-agent systems may prove more resilient and efficient under uncertainty or partial observability.

\section{A Formal (Toy) Example of Drift due to Synthetic Data}\label{formal}

To illustrate the dynamics that may emerge when synthetic data increasingly replaces human-generated language data in the training of large-scale models, let us consider a simple (yet already revealing) formal setup. 

\subsection{The One-Agent Case}

Let \( D^{(H)} \) denote a fixed distribution of human-generated language data, and let \( D_t^{(S)} \) denote the synthetic data distribution produced by a language model \( M_t \) at training step \( t \). The overall training distribution at step \( t \) can be written as a convex combination of the two:

\begin{equation}
D_t = \alpha_t D^{(H)} + (1 - \alpha_t) D_t^{(S)},
\end{equation}

where \( \alpha_t \in [0, 1] \) represents the proportion of human-generated data at step \( t \). It is reasonable to assume that this proportion decreases over time, as high-quality human data becomes scarcer and synthetic data is used more heavily:

\begin{equation}
\frac{d\alpha_t}{dt} < 0.
\end{equation}

The model \( M_t \) itself is updated by a training operator \(\mathcal{T}\), which optimizes a standard objective (for instance, cross-entropy loss) over the current training distribution:

\begin{equation}
M_t = \mathcal{T}(D_t).
\end{equation}

The recursive nature of the process is captured by the fact that the synthetic data at time \(t+1\) is generated by the current model:

\begin{equation}
D_{t+1}^{(S)} = \mathcal{G}(M_t),
\end{equation}

so that

\begin{equation}
D_{t+1} = \alpha_{t+1} D^{(H)} + (1 - \alpha_{t+1}) \mathcal{G}(M_t), \quad M_{t+1} = \mathcal{T}(D_{t+1}).
\end{equation}

In order to reason about the long-term consequences, we introduce a behavioral mapping \(B\) that projects a model \(M\) into a distribution over its observable outputs. We also fix a reference distribution \(B_H\), representing typical human behavior. The divergence between the model’s behavior and human reference at time \(t\) is then given by

\begin{equation}
\delta_t = \mathrm{Dist}(B(M_t), B_H),
\end{equation}

where \(\mathrm{Dist}\) is any suitable statistical divergence (e.g., KL, TV, Wasserstein). Our central hypothesis is that as \(\alpha_t\) declines, synthetic data dominates, and the behavioral divergence \(\delta_t\) grows:

\begin{equation}
\frac{d\delta_t}{dt} > 0.
\end{equation}

This drift is unavoidable unless synthetic data perfectly mimics human data  --  a highly implausible assumption. In the limit, the system may converge to a fixed point \(M^\star\) where training is driven almost entirely by its own outputs, leading to a stable but non-human-like equilibrium:

\begin{equation}
\delta^\star := \mathrm{Dist}(B(M^\star), B_H) > 0.
\end{equation}

This simple one-agent model already conveys the potential hazards of recursive training on synthetic data: the system may slide into a self-referential regime where “humanness” is progressively lost.

\subsection{Extending to Multi-Agent Systems}

The above reasoning assumed a single agent producing and consuming its own outputs. In reality, however, emerging AI ecosystems will be populated by \emph{multiple autonomous agents}, each generating synthetic data and also learning from the outputs of others. This setting is more realistic, but also more concerning, because drift can propagate across agents through their interactions.

Let \(\{M_t^{(i)}\}_{i=1}^m\) denote \(m\) agents co-evolving over time. Each agent produces its own synthetic distribution

\begin{equation}
D_t^{(S,i)} = \mathcal{G}(M_t^{(i)}),
\end{equation}

and updates on a mixture of human data and synthetic data drawn from all agents:

\begin{equation}
D_t^{(i)} = \alpha_t^{(i)} D^{(H)} + (1-\alpha_t^{(i)}) \sum_{j=1}^m w_{ij} D_t^{(S,j)}.
\end{equation}

Here, \(W = [w_{ij}]\) is a matrix describing the influence structure between agents: \(w_{ij}\) is the weight agent \(i\) assigns to synthetic data from agent \(j\), and each row sums to one. In words, this equation says: \emph{each agent is a hybrid learner, anchored to human data but simultaneously influenced by the synthetic traces of others, including itself.} The agent then updates via

\begin{equation}
M_{t+1}^{(i)} = \mathcal{T}(D_t^{(i)}).
\end{equation}

We again define behavioral divergence for each agent:

\begin{equation}
\delta_t^{(i)} = \mathrm{Dist}(B(M_t^{(i)}), B_H).
\end{equation}

\paragraph{Two-agent case.}  
For \(m=2\), suppose each agent learns from a convex mixture of its own and the other’s outputs. Writing the mixing matrix as

\begin{equation}
W = \begin{bmatrix} \beta & 1-\beta \\ 1-\beta & \beta \end{bmatrix}, \quad \beta \in [0,1],
\end{equation}

we see that \(\beta\) controls the extent of self-reliance. If \(\beta\) is high, each agent mainly amplifies its own drift (as in the one-agent case). If \(\beta\) is low, each agent increasingly absorbs the other’s drift. Either way, divergence compounds: one agent’s deviations contaminate the other’s trajectory, and vice versa. Unless a strong human anchor (\(\alpha_t^{(i)}\)) is maintained, both may converge to a coupled but non-human equilibrium.

\paragraph{General \(m\)-agent case.}  
For a network of \(m\) agents, the dynamics are governed by the structure of the weight matrix \(W\). If the influence graph defined by \(W\) is strongly connected (that is, each agent can be indirectly influenced by every other), then any drift introduced by one agent can eventually spread to all. In the extreme case where all \(\alpha_t^{(i)} \to 0\), the system converges to a self-referential regime fully determined by synthetic feedback:

\begin{equation}
M^{(i)\star} = \mathcal{T}\!\Big( \sum_{j=1}^m w_{ij} \, \mathcal{G}(M^{(j)\star}) \Big).
\end{equation}

At such equilibria, the divergence vector \(\boldsymbol\delta^\star = (\delta^{(1)\star}, \dots, \delta^{(m)\star})\) is strictly positive unless all synthetic distributions perfectly mimic human language  --  again, an unrealistic assumption. In other words: \emph{the collective dynamics of interacting synthetic agents do not merely replicate the one-agent drift, but may actually accelerate and entrench it through mutual reinforcement.}

This formal exercise, while admittedly stylized, highlights a crucial point: the risks of synthetic-data drift are not confined to isolated models. In socio-technical systems populated by multiple autonomous agents  --  precisely the scenario we are approaching  --  the recursive use of synthetic data may generate systemic, network-wide deviations from human-like behavior. This possibility, far from being an abstract concern, calls for serious criminological, sociological, and regulatory attention.

\end{document}